%% file: main.tex
\documentclass[10pt,conference]{IEEEtran}

\usepackage{amsmath,amssymb}
\usepackage{graphicx}
\usepackage{booktabs}
\usepackage{xcolor}
\usepackage{hyperref}
\usepackage{cleveref}
\usepackage{balance}
\usepackage{microtype}
\usepackage{multirow}
\usepackage{enumitem}

\hypersetup{colorlinks=true, linkcolor=blue!60!black, citecolor=blue!60!black, urlcolor=blue!60!black}

\begin{document}

\title{When Close Enough Is Not Enough:\\Autoregressive Drift in Quantum Circuit Synthesis}

\author{
\IEEEauthorblockN{Mehdi Saeedi, Eddie Richter, Paul Hartke}
\IEEEauthorblockA{
\textit{Advanced Micro Devices, Inc. (AMD)}\\
\{mehdi.saeedi, eddie.richter, paul.hartke\}@amd.com}
}

\maketitle

% ============================================================
\begin{abstract}
Quantum circuit optimization for fault-tolerant computing requires exact functional equivalence while minimizing expensive non-Clifford resources such as $T$~gates.
We study this problem using a compact 44.8M-parameter encoder-decoder transformer with structured circuit tokenization, evaluating on parameterized circuits (2--6~qubits) and Clifford+T circuits (3--6~qubits).
On parameterized circuits, a hybrid approach---structure from the transformer, angles from classical optimization---achieves median fidelity~1.000 on 3--6~qubit circuits.
On Clifford+T circuits, where all gates are discrete and no post-processing is possible, the model learns valid syntax and accurate $T$-count statistics, yet exact equivalence degrades sharply with target length---from 88\% on circuits with $\leq$9~gates to near zero beyond 26~gates.

We trace this failure to \emph{autoregressive drift}: early-token divergence cascading irrecoverably through left-to-right decoding.
Two levers partially mitigate the drift: inference-time strategies that generate multiple candidates and select via equivalence verification raise exact-match rates from 7\% to 22.5\%, while scaling training data by 2.5$\times$ pushes them to 39.5\%.
Yet the degradation with target length persists---even with more data, exact equivalence drops from 94\% on short circuits to under 4\% beyond 26~gates.

The contrast between settings is our central finding: when approximate outputs can be rescued by post-processing, the transformer succeeds; when exact discrete correctness is required, autoregressive drift limits reliability, with both inference-time search and data scaling as effective levers while training-side fine-tuning and model-level diversification are not.
\end{abstract}


% ============================================================
\section{Introduction}
\label{sec:intro}

Fault-tolerant quantum computing encodes logical qubits into many physical qubits and restricts computation to a set of fault-tolerant operations~\cite{campbell2017roads}.
The standard universal gate set for such architectures is Clifford+T: Clifford gates ($H$, $S$, CNOT) can be implemented transversally at negligible cost, while each non-Clifford $T$~gate requires expensive magic state distillation~\cite{bravyi2005magic}.
Minimizing the \emph{$T$-count}---the total number of $T$ and $T^\dagger$ gates---directly reduces the dominant resource cost of fault-tolerant execution.
In realistic algorithms, $T$-count dominates error-correction overhead, making its optimization a critical compilation step.

Our goal is to test whether a learned model can perform $T$-count optimization, and to pinpoint where it succeeds and fails. Our finding is a sharp dichotomy: pairing a learned structure predictor with classical angle optimization attains near-perfect fidelity on parameterized circuits, whereas for discrete Clifford+T circuits exact equivalence is limited by a failure mode we name \emph{autoregressive drift}---early-token errors that cascade irrecoverably. We state this result and its practical consequences up front, then substantiate it in \Cref{sec:results,sec:analysis}.

Classical $T$-count optimizers employ algebraic rewriting rules derived from ZX-calculus~\cite{pyzx2020}, matroid partitioning~\cite{amy2014tcount}, and phase polynomial representations~\cite{nam2018topt, heyfron2018tcount}.
Circuit simplification via peephole optimization, level compaction~\cite{maslov2008quantum}, and template matching~\cite{maslov2005templates} further reduces gate counts.
For single-qubit unitaries, optimal Clifford+T decomposition is well understood~\cite{kliuchnikov2013synthesis, ross2016optimal}, but multi-qubit synthesis remains challenging.
More broadly, quantum circuit synthesis encompasses a rich set of techniques, from reversible-logic methods~\cite{miller2003transformation, saeedi2013survey} to general unitary synthesis~\cite{shende2006synthesis}.
While effective, they are limited to optimization patterns captured by hand-crafted rules~\cite{arabzadeh2010rule, soeken2010window}; discovering new identities becomes increasingly difficult as circuit complexity grows.

We ask whether a neural network can \emph{learn} circuit optimization rules directly from data,\footnote{We use ``optimization'' for the circuit-to-circuit task studied here---rewriting a given circuit into an equivalent, lower-cost one---a special case of the broader circuit \emph{synthesis} problem named in the title.} potentially replacing or augmenting hand-crafted rule libraries.
To investigate, we train a compact encoder-decoder transformer (44.8M parameters) with a structured tokenization scheme in which every token carries circuit-semantic information, using a vocabulary of 102~tokens.\footnote{4~special tokens (PAD, BOS, EOS, SEP) + 21~gate types + 12~qubit tokens (Q0--Q11) + 1~angle placeholder = 38~base tokens, plus 64~length-conditioning tokens (LEN0--LEN63) = 102 total. For Clifford+T, only 25 of these are active (4~special + 9~CT gates + 12~qubits).}
The architecture handles two complementary tasks:

\begin{itemize}[leftmargin=*]
    \item Parameterized circuits: the model predicts circuit structure with 100\% structural validity across all qubit counts, while a separate regression head outputs $(\sin, \cos)$ angle estimates on the unit circle. Because the predicted structure is near-perfect, angle optimization reduces to a low-dimensional classical problem: a Nelder-Mead optimizer refines the angles, yielding median fidelity~1.000 on 3--6~qubit circuits under beam search, with performance improving at higher qubit counts.
    \item Clifford+T circuits: all gates are discrete and no angle post-processing is possible.
    Trained on 200K PyZX-optimized pairs, the model produces 100\% structurally valid outputs whose average $T$-count matches the PyZX target, yet exact functional equivalence remains limited.
\end{itemize}

The contrast between these settings reveals the core challenge.
On Clifford+T circuits, exact-match rates drop from 88\% on short targets ($\leq$9~gates) to near zero on longer ones ($\geq$26~gates), driven by a failure mode we term \emph{autoregressive drift}: the model diverges from the correct output within the first few tokens, and left-to-right decoding propagates the error through the entire sequence.
On parameterized circuits, similar early-sequence errors arise in angle predictions, but classical post-processing \emph{rescues} the output because the underlying structure is correct.
This isolates the bottleneck: not the transformer's ability to learn optimization patterns, but the impossibility of recovering from discrete token errors once they occur.
From this analysis, we make four contributions:
\begin{enumerate}[leftmargin=*]
    \item A unified transformer architecture with structured tokenization achieving median fidelity~1.000 on 3--6~qubit parameterized circuits (via angle post-processing) and 100\% structural validity on Clifford+T circuits (\Cref{sec:param_results}).
    \item A reproducible diagnosis of autoregressive drift in exact discrete circuit synthesis, identifying target circuit length---not qubit count---as the dominant predictor of failure (\Cref{sec:analysis}).
    \item Inference-time decoding strategies that raise Clifford+T exact-match rates from 7\% (greedy) to 22.5\% (Temp $N{=}200$), with explicit quality--latency tradeoffs (\Cref{sec:results}); the model's high reliability on short circuits suggests deployment as a peephole-style optimization pass within production compilers~(\Cref{sec:discussion}).
    \item Evidence that targeted fine-tuning, candidate pooling, and model ensembling each yield only marginal gains over the 21.0\% Temp $N{=}100$ baseline (M1: 22.5\%, M2: 20.0\%, M3: 21.5\%, M4: 24.5\%), while 2.5$\times$ more training data nearly doubles exact-match rates---yet preserves the drift-driven failure pattern (\Cref{sec:negative}).
\end{enumerate}

% ============================================================
\section{Related Work}
\label{sec:related}

\subsection{Classical T-Count Optimization}

The ZX-calculus provides a graphical language for reasoning about quantum circuits; PyZX~\cite{pyzx2020} leverages it to achieve substantial $T$-count reductions through automated simplification.
Amy et~al.~\cite{amy2014tcount} formulate $T$-depth optimization as a matroid partitioning problem with polynomial-time algorithms for a restricted circuit class, while Nam et~al.~\cite{nam2018topt} combine phase-polynomial rotation merging with gate-cancellation heuristics (Hadamard reduction, CNOT cancellation).
Heyfron and Campbell~\cite{heyfron2018tcount} introduce a compiler applying a sequence of $T$-reducing transformations, and Amy et~al.~\cite{amy2013meet} give a meet-in-the-middle algorithm for depth-optimal Clifford+T synthesis of small unitaries.
Architecture-aware compilation further constrains the problem: mapping circuits to hardware with limited qubit connectivity~\cite{zulehner2019mapping} introduces additional gates that compound the $T$-count overhead.
These approaches are exact and efficient but rely on manually discovered identities, limiting applicability to patterns human experts have already identified.

\subsection{ML for Quantum Circuit Compilation}

\textbf{Prompt-based LLMs.}
QUASAR~\cite{quasar2025} augments a general-purpose language model with tool use to generate quantum assembly (QASM~\cite{cross2022openqasm}) as raw text.
While this can produce valid circuits, it devotes billions of parameters to parsing syntactic tokens---brackets, semicolons, whitespace---that carry no circuit-semantic information.

\textbf{Diffusion models.}
genQC~\cite{genqc2024} encodes circuits as structured gate-position tensors and generates them via denoising diffusion.
The approach produces diverse candidates but struggles with exact equivalence---denoising introduces rounding errors, and fixed-size tensors handle variable-length circuits poorly.

\textbf{Reinforcement learning.}
Rietsch et~al.~\cite{rietsch2024alphazero} apply Gumbel AlphaZero to Clifford+T unitary synthesis, using Monte Carlo tree search (MCTS) guided by a learned value function.
IBM's RL approach~\cite{ibmrl2024} trains agents to synthesize Linear Function, Clifford, and Permutation circuits subject to device connectivity constraints, using step-by-step gate construction with correctness-based rewards.
AlphaTensor-Quantum~\cite{alphatensor_quantum2024} casts $T$-count optimization as a tensor-decomposition problem solved by deep RL with MCTS, avoiding autoregressive decoding entirely.
These methods achieve strong results on specific circuit classes but require custom reward engineering and tend to be sample-inefficient.

\textbf{Generative flow networks.}
FlowQ-Net~\cite{flowqnet2025} learns a stochastic policy that samples VQA ansatz architectures in proportion to a user-defined reward, explicitly targeting diversity rather than RL-style single-trajectory optimization.
Like AlphaZero-based approaches, it avoids autoregressive left-to-right decoding, but it targets ansatz design for variational algorithms (VQE, MaxCut) rather than $T$-count optimization.

\textbf{Search-based and formal methods.}
Quartz~\cite{xu2022quartz} auto-generates a verified database of equivalent circuit classes (ECCs) via SMT solving and applies cost-based backtracking search over this database.
Paradis et~al.~\cite{paradis2024synthetiq} use simulated annealing with a domain-specific energy function and a custom simplification pass to synthesize circuits over arbitrary finite gate sets.
These guarantee correctness but face combinatorial explosion on larger circuits.

\textbf{Evolutionary methods.}
Genetic algorithms have long been applied to circuit compilation and optimization, evolving a population of candidate circuits under fitness functions for depth, gate count, or fidelity~\cite{rasconi2019ga, arufe2022ga}, with hybrids that add a local-search step (e.g., Nelder-Mead) to refine variational parameters~\cite{wakaura2021gals}.
Like the reinforcement-learning and search methods above, these explore many candidates \emph{per instance} rather than amortizing optimization into a single learned autoregressive pass.

\textbf{Supervised learning.}
Thei{\ss}inger et~al.~\cite{theissinger2026beyond} train a lightweight MDL-prediction model that estimates the remaining gate cost of a residual unitary, then use this estimate as a value function inside a stochastic beam search over Clifford+T gates---the closest prior work to ours.
Key differences remain: their input is the target unitary (not an unoptimized input circuit), the supervised target is a scalar MDL (not a token sequence), and they do not diagnose or analyze autoregressive drift.

\textbf{Test-time compute scaling.}
Snell et~al.~\cite{snell2024scaling} show that scaling inference-time compute (e.g., generating and verifying multiple candidates) can outperform scaling model parameters for reasoning tasks.
We instantiate this finding in quantum compilation, where an efficient verifier (unitary matrix comparison) makes generate-and-verify strategies practical.

% ============================================================
\section{Problem Setup}
\label{sec:setup}

\subsection{Task Definition}

We formulate Clifford+T optimization as sequence-to-sequence translation.
Given an input circuit $C_\text{in}$ on 3--6~qubits over the gate set $\mathcal{G} = \{H, S, S^\dagger, T, T^\dagger, \text{CNOT}, \text{CZ}, X, Z\}$ (CZ appears only in PyZX-optimized targets; cf.~\Cref{sec:datagen}), the goal is to produce an output circuit $C_\text{out}$ over the same gate set satisfying:
\begin{enumerate}[leftmargin=*]
    \item Exact equivalence up to a global phase: process fidelity $F(U_{C_\text{in}}, U_{C_\text{out}}) > 0.999$.
    \item $T$-count reduction: the number of $T$ and $T^\dagger$ gates in $C_\text{out}$ is minimized.
\end{enumerate}

Unlike unitary synthesis---where the input is a matrix---the model receives a \emph{circuit} and must learn to recognize and eliminate redundancy in the gate sequence.
This circuit-to-circuit setting enables structured tokenization (\Cref{sec:tokenization}) and provides richer input features than a raw unitary matrix.

\subsection{Metrics}

We use process fidelity as the primary correctness metric.
Given unitary matrices $U$ (input circuit) and $V$ (predicted circuit), we compute:
\begin{equation}
    F(U, V) = \frac{|\mathrm{Tr}(U^\dagger V)|^2}{d^2},
    \label{eq:fidelity}
\end{equation}
where $d = 2^n$ is the Hilbert space dimension for $n$ qubits.
$F = 1.0$ indicates identical operations up to a global phase; we define exact equivalence as $F > 0.999$, with the tolerance accounting for floating-point arithmetic.
For $n \leq 6$ qubits, the unitary is at most $64 \times 64$, making explicit computation of \eqref{eq:fidelity} tractable and enabling our generate-and-verify strategy.

To optimize non-Clifford resource cost, we treat $T$-count as the primary objective and report both predicted and PyZX target $T$-counts to assess whether the model has learned the correct compression level.
Inference latency is the wall-clock time per circuit, including all candidate generation and verification steps.

\subsection{Evaluation Protocol}

All evaluations use a fixed test set of 500 Clifford+T circuits (125 per qubit count, 3--6~qubits), generated independently from the training data with the same random generation pipeline.
Results are stratified by target gate-count bins (1--9, 10--14, 15--25, 26+) to expose the length-dependent failure mode.
All runs use a fixed random seed (42) for reproducibility.
All strategies are evaluated on a 200-sample subset of the test set.

% ============================================================
\section{Method}
\label{sec:method}

\begin{figure}[t]
    \centering
    \includegraphics[width=\columnwidth]{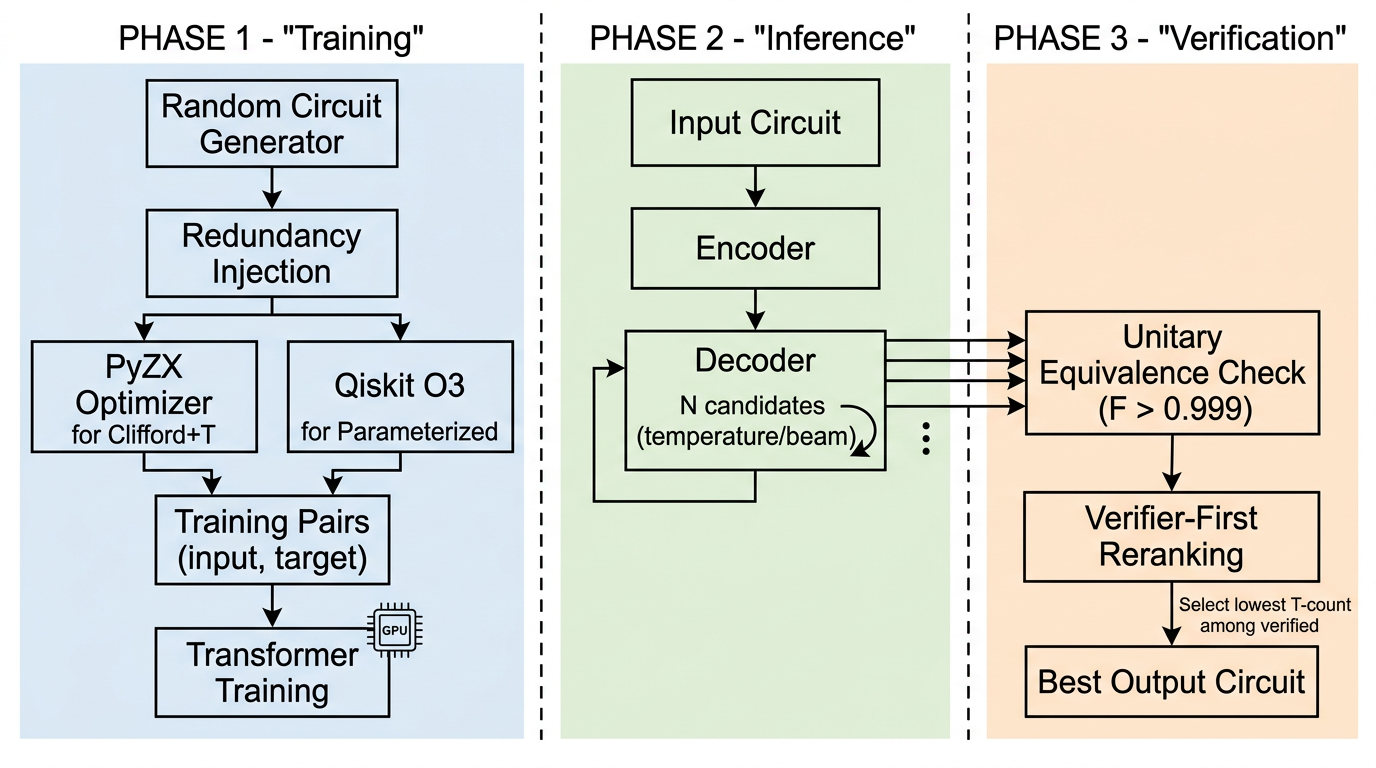}
    \caption{End-to-end pipeline. Training pairs are generated using PyZX (Clifford+T) or Qiskit O3 (parameterized). At inference, $N$ candidates are decoded and verified against the input unitary; the verifier selects the lowest-$T$-count equivalent circuit.}
    \label{fig:pipeline}
\end{figure}

\subsection{Structured Circuit Tokenization}
\label{sec:tokenization}

\Cref{fig:pipeline} illustrates the end-to-end pipeline.
We represent quantum circuits as typed token sequences rather than raw text.
Each gate $g$ on qubits $q_1, \ldots, q_k$ becomes the subsequence $[\texttt{GATE}_g, \texttt{Q}_{q_1}, \ldots, \texttt{Q}_{q_k}]$, and a full circuit is wrapped with delimiters: $[\texttt{BOS}, g_1, \ldots, g_m, \texttt{EOS}]$.

\textbf{Example.}
The 3-qubit circuit $H(q_0) \cdot \text{CNOT}(q_0, q_1) \cdot T(q_2)$ is tokenized as:
\begin{equation*}
[\texttt{BOS},\; \texttt{H},\texttt{Q0},\; \texttt{CX},\texttt{Q0},\texttt{Q1},\; \texttt{T},\texttt{Q2},\; \texttt{EOS}]
\end{equation*}
This 9-token sequence encodes three gates, and every token carries circuit-semantic information---no model capacity is spent on parsing syntactic delimiters.

For Clifford+T circuits, all gates are discrete and no continuous angle parameters appear.
For Clifford+T, the active vocabulary reduces to 25~tokens: 4~special tokens (\texttt{PAD}~padding, \texttt{BOS}~begin-of-sequence, \texttt{EOS}~end-of-sequence, \texttt{SEP}~input/output separator), 9~gate types (H, S, S$^\dagger$, T, T$^\dagger$, CNOT, CZ, X, Z), and up to 12~qubit tokens (Q0--Q11). The remaining 77 tokens (12 additional gate types for parameterized circuits, 1~angle placeholder, and 64~length-conditioning tokens) are unused during Clifford+T inference.
This contrasts with text-based approaches~\cite{quasar2025} that serialize circuits as QASM strings, requiring the model to parse brackets, semicolons, register declarations, and numerical literals; our vocabulary is $\sim$100$\times$ smaller than a typical byte-pair encoding (BPE) vocabulary, and every token directly encodes a circuit operation.

\subsection{Model Architecture}

\begin{figure}[t]
    \centering
    \includegraphics[width=\columnwidth]{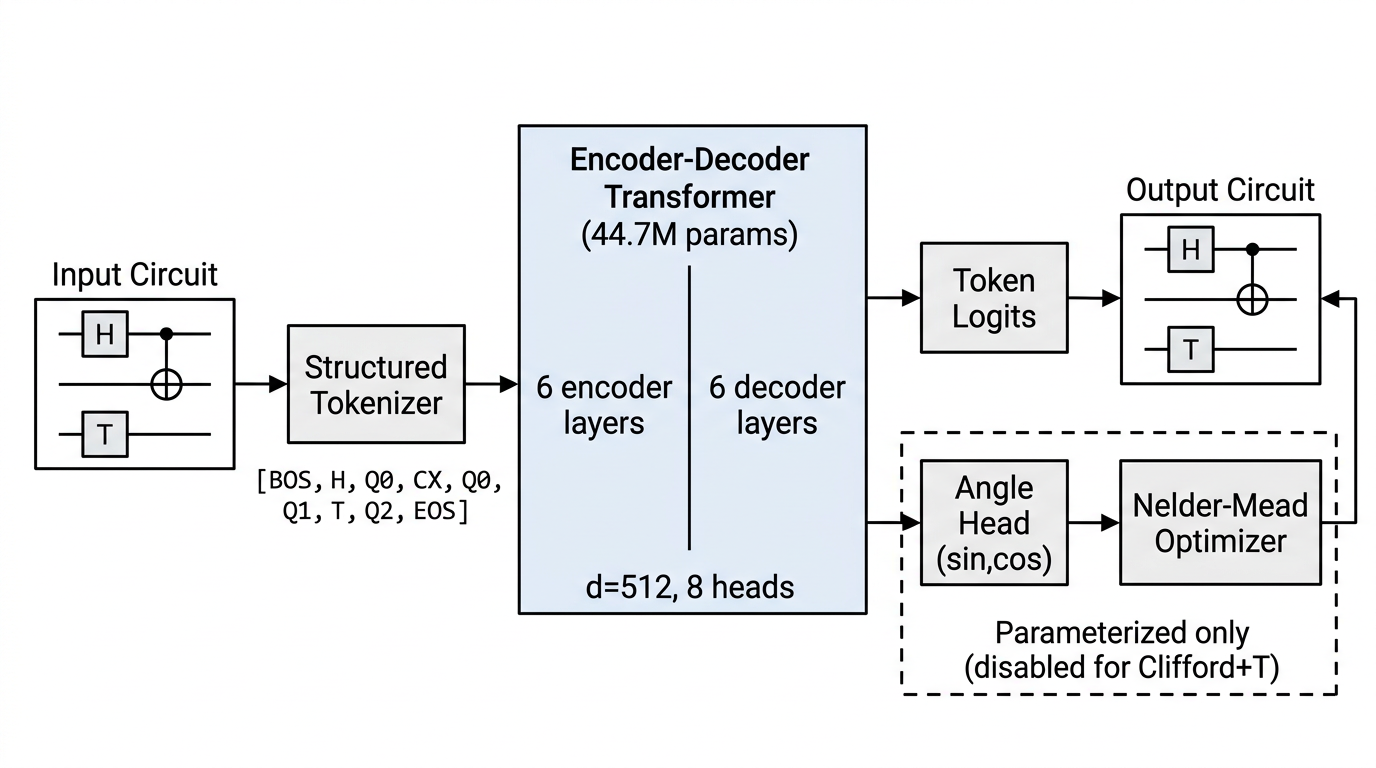}
    \caption{Model architecture. The encoder-decoder transformer produces token logits for circuit structure. For parameterized circuits, a separate angle head predicts $(\sin, \cos)$ pairs refined by Nelder-Mead optimization; this branch is disabled for Clifford+T.}
    \label{fig:architecture}
\end{figure}

We use an encoder-decoder transformer~\cite{vaswani2017attention} with pre-layer normalization (\Cref{fig:architecture}).
The architecture consists of:

\begin{itemize}[leftmargin=*]
    \item Embedding dimension $d_\text{model} = 512$, 8 attention heads
    \item 6~encoder layers and 6~decoder layers
    \item Feedforward dimension $d_\text{ff} = 2048$
    \item Learned token embeddings and learned positional embeddings (max length 256)
    \item Dropout rate 0.1
    \item Xavier uniform initialization
\end{itemize}

This configuration follows the Transformer-base layout of Vaswani et~al.~\cite{vaswani2017attention}, yielding 44.8M trainable parameters---deliberately compact to enable practical inference latency on client-class GPUs (e.g., 0.2~s greedy decoding per circuit) and reproducibility on a single accelerator; model scaling is discussed in \Cref{sec:limitations}.
The decoder uses causal masking---each output token attends only to previously generated tokens and the full encoder output---and a linear projection maps hidden states to logits over the vocabulary.

For parameterized circuits containing rotation gates ($R_Z$, $R_X$, $R_Y$), the model uses a separate angle regression head to predict continuous rotation angles.
Because angles are cyclic ($0$ and $2\pi$ represent the same rotation), naive scalar regression suffers from a discontinuity at the wrap-around.
We resolve this by predicting $(\sin\theta, \cos\theta)$ pairs on the unit circle, recovering the angle via $\text{atan2}(\sin\theta, \cos\theta)$.

The angle head is a 3-layer MLP ($d_\text{model} \to d_\text{model} \to d_\text{model}/2 \to 2$) with GELU activations; angle values are injected into token embeddings at \texttt{ANGLE} positions via a learned projection of the $(\sin, \cos)$ pair, enabling the model to condition on input angles during encoding.
For Clifford+T circuits, all gates are discrete and the angle head is disabled ($\lambda_\text{angle} = 0$).

A key finding is that the transformer learns circuit \emph{structure} almost perfectly---100\% validity, correct gate counts---but produces imprecise angle predictions (validation MAE from 77$^\circ$ on 2-qubit circuits to 6.4$^\circ$ on 6-qubit circuits, where greater structural regularity aids prediction).
For angle post-processing, we exploit this asymmetry by decoupling the two sub-problems: the transformer predicts circuit structure, then Nelder-Mead optimization searches for angles that maximize fidelity with the target unitary.
Post-processing takes approximately 1--2~seconds per circuit and substantially improves average fidelity (e.g., from 0.07 to 0.85 after optimization on 2-qubit circuits with beam search).
Crucially, this decoupling is unavailable for Clifford+T circuits, where every gate is discrete and a single incorrect token changes the output unitary entirely---an asymmetry central to our analysis of autoregressive drift in \Cref{sec:analysis}.

The training objective differs by task. For Clifford+T circuits, the model minimizes cross-entropy loss on the target token sequence:
\begin{equation}
    \mathcal{L}_\text{CT} = -\frac{1}{T} \sum_{t=1}^{T} \log p_\theta(y_t \mid y_{<t}, x)
\end{equation}
For parameterized circuits, the loss combines cross-entropy with angle regression:
\begin{equation}
    \mathcal{L}_\text{param} = \mathcal{L}_\text{CE} + \lambda_\text{angle} \cdot \mathcal{L}_\text{angle}
\end{equation}
where $\mathcal{L}_\text{angle}$ is MSE between predicted and target $(\sin, \cos)$ pairs at \texttt{ANGLE} positions.

\subsection{Training Data Generation}
\label{sec:datagen}

\textbf{Parameterized circuits.}
We generate 200K training pairs per qubit count (each trained as a separate model), consisting of a noisy input circuit and a Qiskit O3-optimized target.
Input circuits contain 5--25 random gates sampled from a broad gate set ($\{H, S, T, X, \ldots\}$ for 1-qubit, $\{R_Z, R_X, R_Y\}$ for rotations, and CX); noise is injected through random gate insertions and identity decompositions, and Qiskit's highest optimization level (O3) transpiles targets to the basis $\{$CX, $R_Z$, SX, $X\}$.
Each example provides a clear training signal from input to O3-optimized target, with the target consistently shorter in both total gate count and CX count.

\textbf{Clifford+T circuits.}
We generate 200K training pairs of (unoptimized input, PyZX-optimized target) across 3--6~qubits (50K per qubit count) via the following pipeline:

\begin{enumerate}[leftmargin=*]
    \item \emph{Random circuit generation}: sample circuits with 10--50 Clifford+T gates, drawn uniformly from $\{H, S, S^\dagger, T, T^\dagger, X, Z, \text{CNOT}\}$ with qubits sampled uniformly.
    \item \emph{Redundancy injection}: insert identity-equivalent gate pairs at random positions---$HH = I$, $SS^\dagger = I$, and $\text{CNOT} \cdot \text{CNOT} = I$---creating suboptimal circuits with known reducible patterns.
    \item \emph{PyZX optimization}: apply \texttt{full\_reduce} (ZX-calculus simplification) followed by \texttt{basic\_optimization} (template matching and phase merging).
    \item \emph{Phase decomposition}: PyZX outputs rotation gates $R_Z(k\pi/4)$; we decompose these to discrete gates via $R_Z(\pi/4) \mapsto T$, $R_Z(\pi/2) \mapsto S$, $R_Z(\pi) \mapsto Z$, and their inverses. PyZX may also introduce CZ gates not present in the input, extending the target gate set to $\mathcal{G}$.
    \item \emph{Verification}: confirm unitary equivalence ($F > 0.999$) and positive $T$-count reduction; discard failing pairs ($<0.1\%$).
\end{enumerate}

The resulting dataset has an average $T$-count reduction of $\sim$65\% (input avg.\ 8.10 $\to$ target avg.\ 2.83 $T$-gates; cf.\ \Cref{tab:tcount}).
Validation (3K) and test (500) sets are generated independently from the training data.

\subsection{Decoding Strategies}
\label{sec:decoding}

We evaluate five decoding strategies, all sharing a \emph{verifier-first reranking} policy for candidate selection.
\emph{(1) Greedy decoding} selects the highest-probability token at each step ($N{=}1$).
\emph{(2) Beam search} maintains $w{=}5$ partial hypotheses scored by length-normalized log-probability and returns the top-$w$ complete sequences.
\emph{(3) Temperature sampling} generates $N$ independent candidates by sampling at $\tau{=}0.8$ with nucleus filtering~\cite{holtzman2020nucleus} (top-$p{=}0.95$), providing stochastic diversity.
\emph{(4) Length-banded sampling} generates candidates conditioned on different target lengths (predicted length $\pm$4 gates), spreading the budget across multiple length hypotheses; we evaluate this strategy because 87\% of failures involve incorrect output length, although the empirical results (\Cref{sec:results}) show it underperforms plain temperature at the same budget.
\emph{(5) Hybrid search} pools candidates from multiple temperatures ($\tau \in \{0.75, 0.85, 0.95\}$), both length-banded and non-banded, then applies verifier-first reranking over the merged pool.

All candidate-based strategies use verifier-first reranking: compute fidelity $F$ between each candidate and the input unitary, select those with $F > 0.999$, and return the lowest-$T$-count verified candidate---or the highest-fidelity one if none passes.
Correctness takes priority over compression.

% ============================================================
\section{Experimental Setup}
\label{sec:experiments}

All experiments run on AMD Instinct\texttrademark{} MI300X GPUs (192~GB HBM3). For training, the Clifford+T model trains for 40~epochs with batch size~48 using AdamW ($\text{lr} = 3 \times 10^{-4}$, $\beta_1 = 0.9$, $\beta_2 = 0.999$, weight decay $10^{-2}$) and a cosine schedule with 500-step linear warmup.
Training completes in approximately 3.7~hours.
Each parameterized model (one per qubit count) trains for 60~epochs with the same batch size and learning rate, completing in approximately 5.5~hours each.
Convergence is smooth across all models, with small train--validation gaps indicating minimal overfitting.

For evaluation, all strategies use $\tau{=}0.8$, top-$p{=}0.95$, and a fixed seed (42).
All strategies---greedy, beam, and candidate-based ($N \in \{20, 50, 100\}$)---are evaluated on 200 test samples.
The largest configuration (hybrid search, $N{=}100$) requires $\sim$21~seconds per circuit.

As a baseline, PyZX~\cite{pyzx2020}~(\texttt{full\_reduce + basic\_optimization}) serves as both the training oracle and the primary baseline, achieving 100\% exact equivalence at approximately 6~ms per circuit.
Our goal is not to outperform PyZX but to study the capabilities and failure modes of learned models on this task.

For reproducibility, each run produces a machine-readable JSON summary with full hyperparameters, per-sample results, and per-length-bin breakdowns under fixed random seeds.

% ============================================================
\section{Results}
\label{sec:results}

\subsection{Parameterized Circuit Optimization}
\label{sec:param_results}

Before turning to Clifford+T---the primary focus---we establish that the architecture is effective on parameterized circuits, where angle post-processing is applicable.
\Cref{tab:param} summarizes results on 200 test samples per qubit count, with separate models trained for each.
Without post-processing, raw fidelity is modest because angle predictions carry substantial error.
However, 100\% of generated circuits are structurally valid with correct gate counts across all qubit counts, confirming that the transformer learns circuit \emph{structure} reliably regardless of problem size.
Holding the predicted structure fixed and optimizing only rotation angles via Nelder-Mead improves performance dramatically:

\begin{itemize}[leftmargin=*]
    %   2q beam: F>0.99=58.5%, avg=0.849, med=0.999
    %   3q beam: F>0.99=81.0%, avg=0.929, med=1.000
    \item 2--3~qubits: angle optimization achieves median fidelity 1.000 for 3q (0.999 for 2q under beam search), with validation token accuracy of 92.2\% (2q) and 97.9\% (3q).
    %   4q beam: F>0.99=85.5%, avg=0.953, med=1.000
    %   5q beam: F>0.99=90.0%, avg=0.969, med=1.000
    %   6q beam: F>0.99=95.0%, avg=0.987, med=1.000
    \item 4--6~qubits: the same architecture scales smoothly, reaching validation token accuracy of 98.6\% (4q), 99.2\% (5q), and 99.3\% (6q) at the best checkpoints, with angle MAE reaching 6.4$^\circ$ for 6-qubit circuits by the end of training.
\end{itemize}

\input{tables/table5_param}

These results reveal two insights.
First, the transformer's strength lies in learning circuit structure, not predicting exact continuous values---once the structure is correct, angle optimization reduces to a low-dimensional problem (5--15 variables for 2q, up to $\sim$30 for 6q) that classical optimizers solve reliably in $\sim$2~seconds per circuit.
Second, performance \emph{improves} with qubit count: larger circuits provide more structural regularity for the model to exploit, and the angle regression head benefits from the richer context.
This sets up the critical contrast with Clifford+T circuits: when all gates are discrete, there is no ``close enough'' regime, and a single wrong token cannot be rescued by post-processing.

\textbf{Unified vs.\ per-qubit models.}
\Cref{tab:unified} compares the per-qubit models (200K training samples each) against a single unified model trained on 200K samples across all qubit counts (40K per qubit---$5\times$ less per-qubit data).
The unified model \emph{outperforms} per-qubit models on 2--4~qubits (e.g., 67.0\% vs.\ 58.5\% $F{>}0.99$ on 2q beam), suggesting that cross-qubit transfer learning compensates for reduced per-qubit data.
At 5--6~qubits, per-qubit specialization dominates (95.0\% vs.\ 83.5\% on 6q), indicating that higher-qubit circuits benefit from dedicated training data.
This crossover suggests a practical strategy: a unified model for small circuits with per-qubit fine-tuning for larger ones.

\input{tables/table6_unified}

\subsection{Clifford+T: Main Comparison}

\Cref{tab:main} presents the full decoding strategy comparison; four trends emerge.

\emph{Inference-time compute scaling improves exact equivalence.}
Greedy decoding achieves 7.0\% exact equivalence on 200 test samples.
Beam search ($w{=}5$) raises this to 14.0\%---doubling the success rate.
Temperature sampling with $N{=}100$ candidates and verifier-first reranking reaches 21.0\%, while hybrid search (pooling diverse strategies) achieves 21.5\%---a $>$3$\times$ improvement over greedy.
Scaling plain temperature beyond $N{=}100$ yields diminishing returns: Temp $N{=}200$ reaches only 22.5\% (\Cref{tab:main}), adding 1.5~pp over Temp $N{=}100$ at $>$2$\times$ the latency (45~s vs.\ 21~s)---the same pass@$k$ saturation pattern observed in code generation~\cite{brown2024monkeys}.

\emph{Length conditioning hurts; pooled diversity barely helps.}
Despite 87\% of failures involving wrong output length, length-banded sampling \emph{underperforms} plain temperature at every budget ($N{=}100$: 18.0\% vs.\ 21.0\%).
Hybrid pooling across multiple temperatures and length conditions reaches 21.5\% at $N{=}100$, but the 0.5~pp gain over plain temperature lies within its Wilson interval; the dominant lever is raising $N$, not diversifying the sampling distribution.

\emph{Beam search saturates around $w{=}10$.}
A sweep over beam widths $w \in \{1, 2, 3, 5, 10, 20\}$ on 200 test samples shows monotonic improvement with diminishing marginal returns above $w{=}10$: 7.0\% ($w{=}1$, greedy) $\to$ 14.0\% ($w{=}5$) $\to$ 18.0\% ($w{=}10$) $\to$ 19.5\% ($w{=}20$), capturing roughly half of the $>$3$\times$ gain achievable by temperature sampling at $N{=}100$.
The plateau reflects beam search exploring a narrow high-probability region rather than the diverse candidates needed to overcome early-token errors.

\emph{The model's distributional knowledge is accurate.}
Across all strategies, predicted $T$-count closely matches the PyZX target---2.89 vs.\ 2.83 for greedy on 200 samples (\Cref{tab:tcount})---and every generated circuit is a structurally valid Clifford+T sequence.
The model has learned \emph{what optimized circuits look like} in aggregate, even when it fails to produce the correct circuit for a given input.

\input{tables/table1_main}

\subsection{Quality--Latency Tradeoff}

\begin{figure}[t]
    \centering
    \includegraphics[width=\columnwidth]{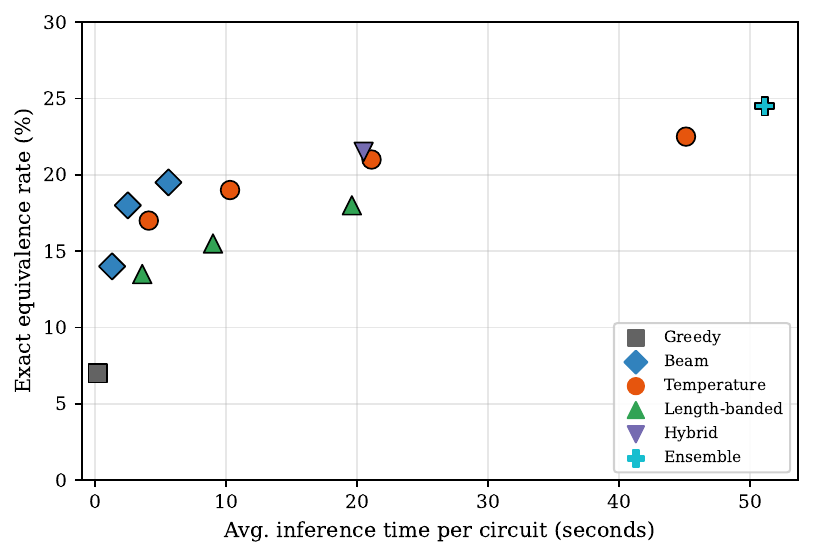}
    \caption{Quality--latency Pareto frontier for Clifford+T optimization. Each point represents a (strategy, candidate budget) configuration evaluated on 200 test samples. Inference-time compute scaling monotonically improves exact equivalence for temperature sampling.}
    \label{fig:pareto}
\end{figure}

\Cref{fig:pareto} shows the quality--latency Pareto frontier. Two operating points emerge: temperature $N{=}50$ (19\% exact, $\sim$10~s/circuit) for throughput-oriented batch use, and hybrid search $N{=}100$ (21.5\%, $\sim$21~s/circuit) when correctness is prioritized; greedy ($\sim$0.2~s) and beam ($\sim$1.3~s) are faster at lower accuracy.
The tradeoff spans two orders of magnitude in latency (0.2--21~s) for a $>$3$\times$ improvement in exact equivalence (7\% to 21.5\%).

\subsection{T-Count Prediction Accuracy}

\Cref{tab:tcount} confirms distributional accuracy on 200 test samples: predicted $T$-count (2.89 greedy, 2.89 beam) closely matches the PyZX target (2.83), with a slight tendency toward over-prediction---the model occasionally retains one extra $T$~gate.
Input circuits average 8.10 $T$-gates; the predicted average (2.89) corresponds to $\sim$64\% reduction in aggregate, matching the PyZX target distribution (2.83, $\sim$65\% reduction).
The limitation is not optimization \emph{quality} (comparable to PyZX) but the \emph{reliability} of producing exactly correct circuits.

To quantify the \emph{realized} optimization---not just distributional accuracy---we isolate the subset of outputs verified equivalent to the input unitary ($F{>}0.999$) and measure the reduction they achieve. On this subset the model attains PyZX-quality optimization: greedy reduces the average $T$-count from 5.50 to 1.64 (a 64\% reduction) and total gate count by 70\% (beam: 61\% and 68\%). In 100\% of verified cases the predicted $T$-count \emph{equals} the PyZX target---the model never produces a worse optimization on a circuit it solves, but it also does not discover reductions \emph{beyond} PyZX. Notably, the verified-equivalent rate coincides with the exact-match rate: every circuit the model optimizes correctly matches PyZX's $T$-count. The optimization quality on solved instances is therefore not the bottleneck; the binding constraint is the reliability of producing an exactly correct circuit in the first place, which \Cref{sec:analysis} traces to autoregressive drift.

\input{tables/table4_tcount}

% ============================================================
\section{Failure Analysis}
\label{sec:analysis}

\subsection{Length Is the Dominant Failure Predictor}

\begin{figure}[t]
    \centering
    \includegraphics[width=\columnwidth]{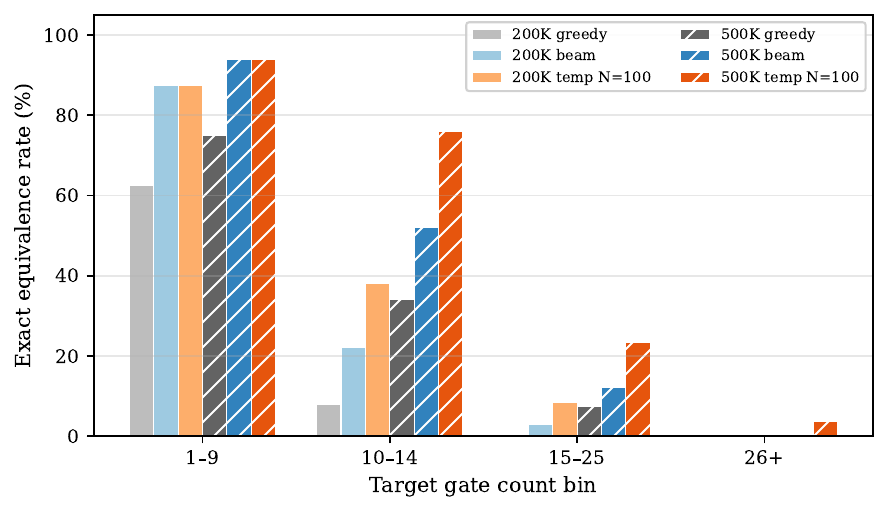}
    \caption{Exact equivalence rate by target gate-count bin for both the 200K baseline (lighter bars) and 500K model (darker, hatched bars). Data scaling improves all length bins substantially, but the degradation pattern persists: performance drops sharply with target length across all configurations.}
    \label{fig:length_bins}
\end{figure}

\Cref{tab:length_bins} and \Cref{fig:length_bins} break down exact-match rates by target circuit length (gates in the PyZX-optimized reference); the pattern is striking:

\begin{itemize}[leftmargin=*]
    \item 1--9 gates: 62.5\% (greedy) to 87.5\% (beam). The model handles short circuits reliably.
    \item 10--14 gates: 8.0\% (greedy) to 44.0\% (hybrid $N{=}100$). Performance drops sharply at the transition from single-digit to double-digit gate counts.
    \item 15--25 gates: 0\% (greedy) to 8.4\% (temperature $N{=}100$). Inference-time scaling provides only marginal improvement.
    \item 26+ gates: 0\% across all strategies. No configuration achieves any exact matches.
\end{itemize}

Critically, \emph{qubit count is a much weaker predictor than length}.
On the full 500-sample test set under greedy decoding, per-qubit exact-match rates span a narrow range (3q: 8.0\%, 4q: 7.2\%, 5q: 15.2\%, 6q: 13.6\%); under the same greedy protocol on the 200-sample subset, length-bin rates range from 62.5\% (1--9~gates) to 0\% (26+~gates), confirming that the dominant bottleneck is sequence length rather than problem dimensionality.

\input{tables/table2_length_bins}

\subsection{Autoregressive Drift: Root Causes}

\begin{figure}[t]
    \centering
    \includegraphics[width=\columnwidth]{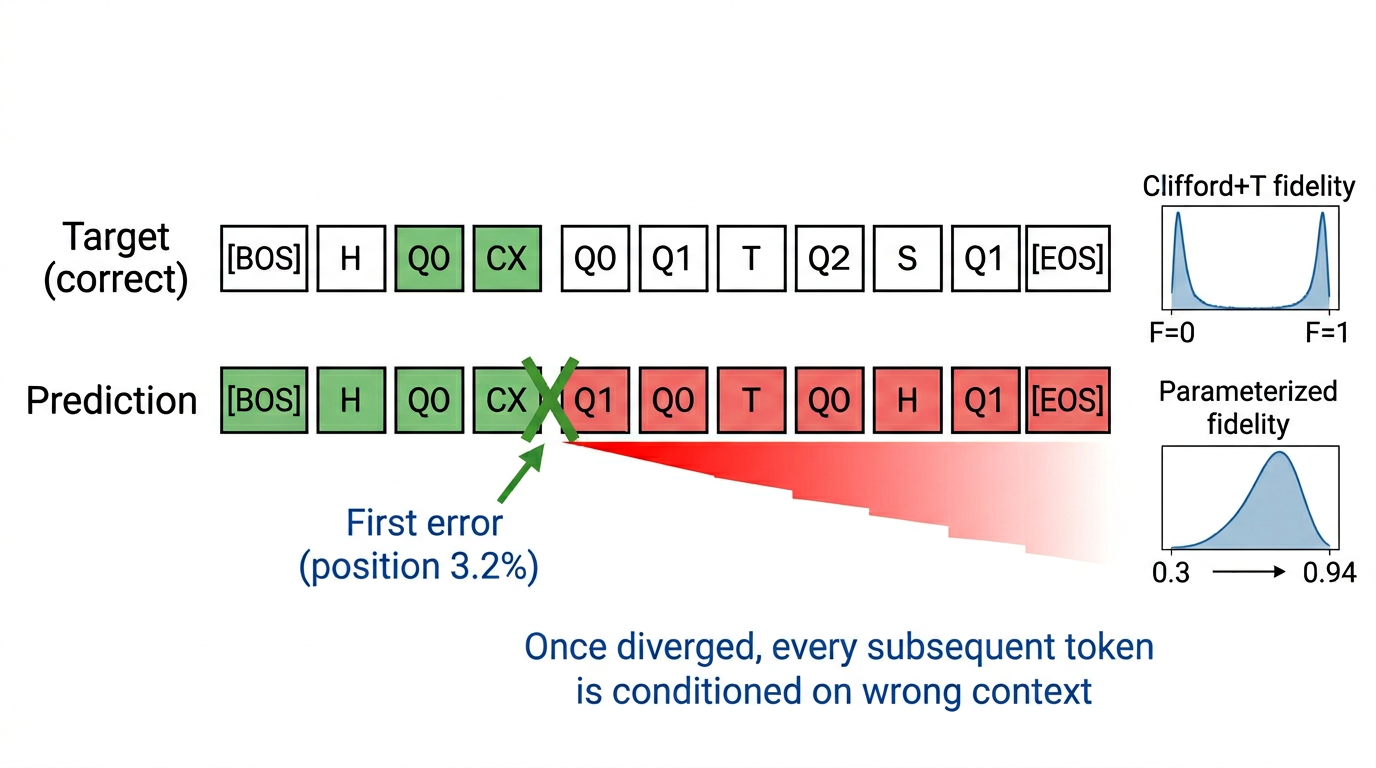}
    \caption{Autoregressive drift illustrated. The model matches the first few tokens (green) but diverges early (red X), after which every subsequent token is conditioned on wrong context. The Clifford+T fidelity distribution is bimodal ($F \approx 0$ or $F \approx 1$), whereas parameterized fidelity is smooth and recoverable via angle post-processing.}
    \label{fig:drift}
\end{figure}

\begin{figure}[t]
    \centering
    \includegraphics[width=\columnwidth]{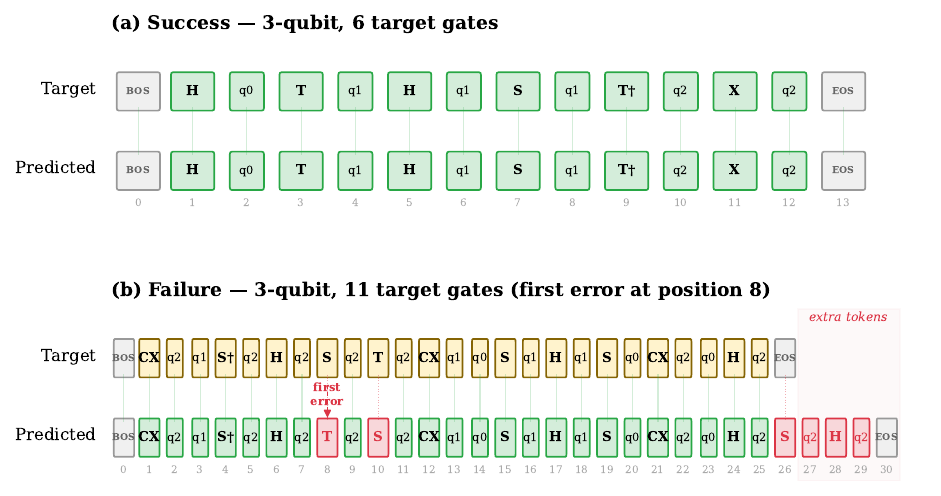}
    \caption{Concrete example from evaluation data. (a)~A short 3-qubit circuit (6~target gates) predicted exactly. (b)~A longer circuit (11~target gates) where the model matches the first 8~tokens but swaps S$\leftrightarrow$T at position~8, triggering cascading errors and extra trailing tokens.}
    \label{fig:example}
\end{figure}

Detailed per-sample diagnostics on the 200-sample evaluation subset reveal three root causes (\Cref{fig:drift}):

\emph{Wrong sequence length.}
87\% of failed predictions produce incorrect-length output, with an average deviation of $\sim$3~gates ($\sim$7~tokens).
Only 13\% of predictions match the target length exactly; however, most length errors are modest (1--5~gates), indicating the model captures approximate output length but diverges in gate-level details.

\begin{figure}[t]
    \centering
    \includegraphics[width=\columnwidth]{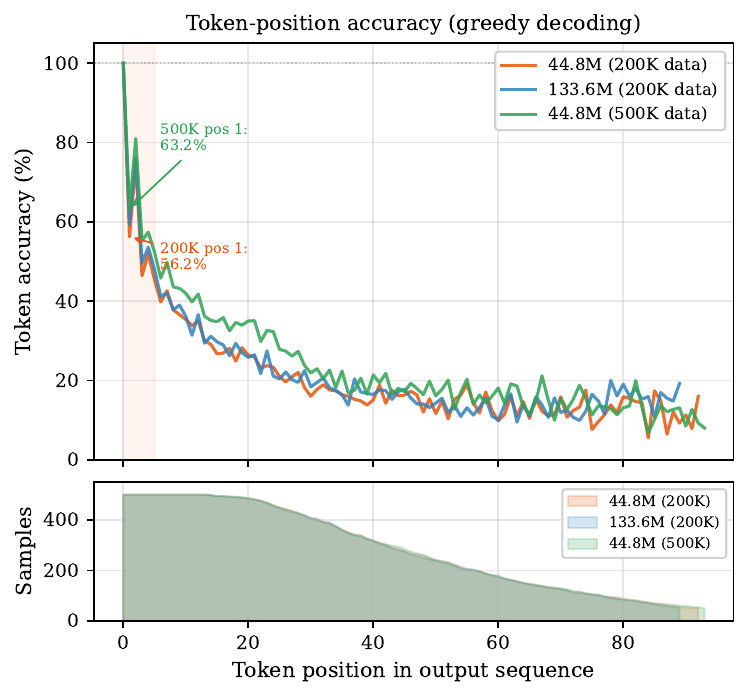}
    \caption{Token-position accuracy under greedy decoding (500 test samples). All three models---44.8M (200K), 133.6M (200K), and 44.8M (500K)---show the same degradation shape. Data scaling (500K) raises the curve uniformly (e.g., 56\% $\to$ 63\% at position~1) but does not change the slope of decline. Neither model scaling (3$\times$ parameters) nor data scaling (2.5$\times$ data) alleviates the drift.}
    \label{fig:position_accuracy}
\end{figure}

\emph{Early divergence.}
The median first-error position is at 3.13\% of the target sequence---within the first 1--2~gates; the 10th percentile is at 1.1\%.
Token-level accuracy analysis on 500 test samples confirms the rapidity of divergence (\Cref{fig:position_accuracy}): the very first content token (position~1, after BOS) is correct only 56.2\% of the time, and accuracy continues to decline, reaching 16.0\% by position~30.
Once an incorrect gate or qubit is emitted, every subsequent token is conditioned on wrong context, producing a cascading error that never recovers (\Cref{fig:example}b).
Crucially, this pattern persists even with 2.5$\times$ training data (500K).
On the 500-sample drift-analysis subset, the 500K model nearly doubles greedy exact match from 11.0\% to 22.2\% and improves first-token accuracy from 56.2\% to 63.2\%.\footnote{These 500-sample drift numbers complement the 200-sample protocol used for \crefrange{tab:main}{tab:data_scaling}; the corresponding 200-sample greedy rates are 7.0\% (200K) and 18.5\% (500K).}
Yet the 500K model's median first-error position (3.57\% of the target sequence) is statistically indistinguishable from the 133.6M model's (3.59\%) and from the 200K baseline's (3.13\%), and all three position-accuracy curves share the same shape (\Cref{fig:position_accuracy}).
Data scaling helps the model ``get it right from the start'' more often, but does not alter the cascading nature of drift once an error occurs.

\emph{Bimodal fidelity distribution.}
Clifford+T fidelity is strongly bimodal: circuits cluster at $F \approx 1.0$ or $F \approx 0$, with virtually nothing between.
This contrasts with parameterized results (\Cref{sec:param_results}), where a smooth fidelity distribution improves substantially via angle post-processing.
The contrast isolates the mechanism: the transformer makes similar early-sequence errors in both settings, but parameterized outputs can be \emph{rescued} by refitting angles, whereas a single wrong discrete gate irreversibly corrupts the unitary.
No post-processing can recover from such errors in the discrete setting.

We term this combination \emph{autoregressive drift}:

\smallskip
\noindent\textbf{Definition.}\enspace Autoregressive drift is a failure mode of left-to-right sequence generation in which an early-token prediction error propagates through the conditioning context, causing all subsequent tokens to diverge from the target---irrecoverably when the output domain is discrete.
\smallskip

\noindent The mechanism explains why scaling $N$ helps: each temperature sample independently avoids (or fails to avoid) the initial divergence, and the verifier selects any successful candidate.
For a circuit with per-sample success probability $p$, the probability of at least one success in $N$ trials is $1 - (1 - p)^N$, which saturates once $Np \gg 1$.
This yields a two-part pattern: (i)~for circuits with $p > 0$, modest $N{=}50$--$100$ drives per-circuit success close to 1; (ii)~the 88\% ceiling on the 1--9~gate bin reflects the fraction of \emph{solvable} short circuits---the $\sim$12\% remainder have $p \approx 0$, and no candidate budget rescues them.
The failure mode is thus not ``almost right'' but complete early divergence.

% ============================================================
\section{Interventions}
\label{sec:negative}

Given the autoregressive drift diagnosis, we tested five targeted interventions spanning training-time fine-tuning, inference-time compute, model diversity, and data scaling (\Cref{tab:negative}).
Each targets a specific hypothesis about the root cause; all evaluations use the same locked protocol and test subset as the baseline.

\textbf{M1: Long-only fine-tuning.}
\emph{Hypothesis}: concentrating training on long circuits will improve the 15--25~gate bin.
\emph{Method}: fine-tune the 200K baseline for 5~epochs on the $\geq$15-gate subset (151{,}328 samples, $\text{lr}=10^{-4}$).
\emph{Result}: at Temp $N{=}100$, 22.5\% exact on 200 test samples (vs.\ 21.0\% baseline), and the targeted 15--25~gate bin is \emph{unchanged} at 8.4\% (9/107).

\textbf{M2: Weighted sampling.}
\emph{Hypothesis}: upweighting long circuits while preserving the full distribution avoids catastrophic forgetting.
\emph{Method}: fine-tune 3~epochs on the full 200K set with $2\times$ oversampling of $\geq$15-gate examples ($\text{lr}=5\times10^{-5}$).
\emph{Result}: 20.0\% exact on 200 samples (vs.\ 21.0\% baseline); 15--25~gates drops to 7.5\% (8/107).

\textbf{M3: Hybrid search.}
\emph{Hypothesis}: pooling candidates from diverse strategies increases the probability of finding a correct one.
\emph{Method}: generate at $\tau \in \{0.75, 0.85, 0.95\}$ from both length-banded and plain sampling, pool $N{=}100$ total, apply verifier-first reranking.
\emph{Result}: 21.5\% exact equivalence on 200 test samples (vs.\ 21.0\% for plain temperature at $N{=}100$).
\emph{Conclusion}: pooling candidates from diverse strategies provides only marginal improvement in \emph{correct}-candidate diversity.

\textbf{M4: Model ensemble.}
\emph{Hypothesis}: pooling candidates from architecturally diverse models yields more diverse correct circuits than a single model at the same budget.
\emph{Method}: generate $N{=}100$ candidates from each of two models (44.8M and 133.6M parameters, 200 total) at $\tau{=}0.8$, pool, apply verifier-first reranking (200 test samples).
\emph{Result}: 24.5\% exact equivalence---above hybrid search (21.5\%) at the same total budget, and marginally above single-model $N{=}200$ temperature sampling (22.5\%).
\emph{Conclusion}: architecturally diverse models do not produce meaningfully more diverse \emph{correct} candidates, suggesting model errors are correlated across architectures trained on the same data.

\textbf{M5: Training data scaling (2.5$\times$).}
\emph{Hypothesis}: the model is data-limited; more training examples will improve per-sample success probability.
\emph{Method}: generate 500K Clifford+T training pairs (2.5$\times$ the original 200K) with the same pipeline (\Cref{sec:datagen}), train the identical 44.8M architecture, and evaluate all strategies on 200 test samples.
\emph{Result}: data scaling produces \emph{significant} gains across all strategies (\Cref{tab:data_scaling}): greedy improves from 7.0\% to 18.5\% ($2.6\times$), beam from 14.0\% to 27.0\% ($1.9\times$), and temperature $N{=}100$ from 21.0\% to 39.5\% ($1.9\times$)---an absolute gain of 11--19~pp at every operating point.
The improvement extends to longer circuits: exact equivalence on the 15--25~gate bin rises from 8.4\% to 23.4\%, and for the first time a non-zero result appears on circuits with 26+ gates (3.7\%, i.e., one out of 27).
Conversely, reducing data below 200K reveals a sharp cliff: models trained on 50K and 100K samples achieve 0\% exact equivalence across all strategies (200 test samples), despite training to convergence.
\emph{Conclusion}: the model is not data-saturated, and more training examples improve the per-sample success probability $p$ that drives both greedy accuracy and the effectiveness of candidate-based search.
A sharp transition occurs between 100K (0\%) and 200K ($\geq$7\%) training pairs; the precise threshold lies in this interval but we did not characterize it further.
Crucially, however, the drift-driven degradation pattern persists: even with 2.5$\times$ data, exact equivalence still drops from 94\% ($\leq$9~gates) to 3.7\% (26+ gates)---a 25$\times$ degradation ratio.
Token-level drift analysis confirms this: the 500K model's median first-error position (3.57\% of the sequence) is comparable to the 200K baseline (3.13\%), and the position-accuracy curve (\Cref{fig:position_accuracy}) shifts upward uniformly without changing slope.

\input{tables/table3_negative}
\input{tables/table7_data_scaling}

\textbf{Summary.} Only data scaling (M5) yields meaningful gains; M1--M4 leave the 15--25~gate bin near the 8.4\% baseline.

\begin{figure}[t]
    \centering
    \includegraphics[width=\columnwidth]{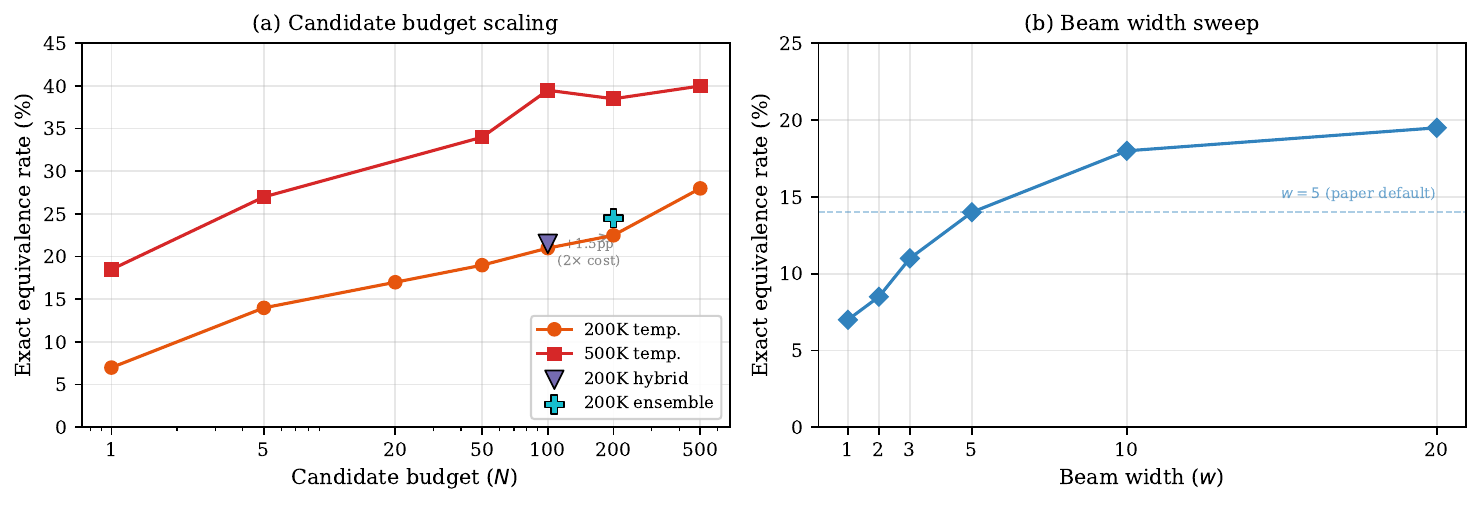}
    \caption{Exact equivalence rate vs.\ candidate budget $N$. Both the 200K baseline and 500K model improve monotonically with more candidates but exhibit strong diminishing returns. Data scaling (2.5$\times$) shifts the entire curve upward, providing larger gains than any inference-time strategy alone.}
    \label{fig:scaling}
\end{figure}

\Cref{fig:scaling} shows the scaling relationship between candidate budget and exact-equivalence rate for both the 200K baseline and the 500K model.
Temperature sampling improves monotonically as $N$ increases but with strong diminishing returns: the 200K model gains only 1.5~percentage points from $N{=}100$ to $N{=}200$ (21.0\% $\to$ 22.5\%), and even $N{=}500$ reaches only 28.0\% on a 50-sample subset.
The 500K model exhibits the same diminishing-returns pattern at a uniformly higher level: $N{=}200$ achieves 38.5\% (vs.\ 39.5\% at $N{=}100$), and $N{=}500$ reaches only 40.0\% (also on a 50-sample subset)---suggesting a practical ceiling of $\sim$40\% for the current architecture and data scale.
Hybrid search achieves the best 200K result at $N{=}100$ (21.5\%), but is surpassed by the 500K model at every budget.

% ============================================================
\section{Discussion}
\label{sec:discussion}

\subsection{Inference-Time vs.\ Training-Time Compute}

For tasks requiring exact correctness with efficient verifiers, scaling inference-time compute provides reliable gains over greedy decoding~\cite{snell2024scaling, chen2021codex}---but scaling training \emph{data} provides larger gains still: the 2.5$\times$ data experiment (M5) nearly doubles exact equivalence at every operating point, while fine-tuning on the same 200K data (M1, M2) is statistically indistinguishable from the baseline.
This suggests that the per-sample success probability $p$ is the fundamental quantity: data scaling increases $p$, amplifying subsequent inference-time search.
The structural requirement is \emph{verification asymmetry}: unitary comparison ($O(4^n)$, $\sim$0.1~ms for 6 qubits) is orders of magnitude faster than autoregressive generation ($\sim$0.2~s per candidate), so the verifier scales freely with $N$.

\subsection{Practical Deployment Considerations}

The model's strength on short circuits (e.g., $\sim$74\% exact equivalence for the 500K model on $\leq$14-gate subcircuits) maps directly to \emph{peephole optimization} in classical compilers.
Production tools like Qiskit~\cite{qiskit2024}, tket~\cite{sivarajah2021pytket}, and MQT~\cite{wille2024mqt} decompose circuits into small subcircuits (2--4 qubits, 5--15 gates) and optimize each independently.
A model with 100\% structural validity and 62--88\% exact equivalence on this range can serve as a complementary optimization pass: invoke it on matching subcircuits, fall back to PyZX on verification failure.

\subsection{Comparison with PyZX}

PyZX achieves 100\% exact equivalence at $\sim$6~ms per circuit---faster and more reliable than our model at every operating point.
We do not position the learned model as a replacement; the contributions are \emph{diagnostic} (identifying autoregressive drift as the bottleneck), \emph{methodological} (establishing inference-time scaling and data scaling as the most effective levers), and \emph{empirical} (a reproducible benchmark with explicit operating points).
The long-term potential of learned models lies in discovering patterns not captured by existing identities; whether the model's learned rewrites extend beyond PyZX's repertoire is best tested on circuit families where PyZX is known to be suboptimal.

\subsection{Relation to Drift in Other Domains}

Autoregressive drift shares the \emph{mechanism} of \emph{exposure bias} in neural machine translation~\cite{bengio2015scheduled, ranzato2016sequence}, but the \emph{consequence} differs: translation errors degrade quality smoothly, whereas a single wrong discrete gate irreversibly corrupts the unitary---the bimodal $F\!\approx\!0$ vs.\ $F\!\approx\!1$ regime of \Cref{sec:analysis}, with no partial credit.
This also explains why pass@$k$ scaling, effective in code generation~\cite{chen2021codex, brown2024monkeys}, saturates earlier here: code training enjoys partial-credit feedback (compile errors, failing tests) that Clifford+T pairs lack, so mitigations such as execution-guided decoding must be reformulated without that gradient-shaping signal.

% ============================================================
\section{Limitations and Future Work}
\label{sec:limitations}

All strategies are evaluated on 200 test samples; the 95\% Wilson confidence interval for a 20\% rate at $n{=}200$ spans $\pm$6 percentage points.

Exact equivalence on circuits with $\geq$15 target gates remains below 9\% for the 200K model; the 500K model pushes this to 23.4\% on 15--25~gates but still reaches only 3.7\% on 26+ gates.
Breaking this ceiling likely requires departing from pure autoregressive decoding---for example, non-autoregressive decoders~\cite{gu2018nonautoregressive} that generate all tokens in parallel with iterative refinement, subcircuit decomposition with a keep/remove gate mask, or peephole windowing that operates within the model's high-accuracy regime.

Our contribution is a diagnosis, not a leaderboard entry, so we compare only against the PyZX oracle; head-to-head comparisons with other ML approaches (genQC~\cite{genqc2024}, RL methods~\cite{ibmrl2024, rietsch2024alphazero}) are confounded by differing gate sets, input representations, and objectives.
Our results are likewise conditioned on the training distribution---random circuits with injected redundancy optimized by PyZX---so generalization to circuits emitted by production compilers remains untested.
The more pointed question---whether non-autoregressive architectures (discrete diffusion, MCTS, iterative refinement) escape the drift failure mode---is a direct target for future work.

\textbf{Model scaling.}
Primary experiments use a 44.8M-parameter model on a single MI300X accelerator.
A 133.6M-parameter variant ($d_\text{model}{=}768$, $h{=}12$, 8+8 layers) improves greedy to 9.5\% and beam to 16.5\% (vs.\ 7.0\%/14.0\% for 44.8M), but ensembling it with the 44.8M model (M4) yields only 24.5\%---within the Wilson interval of single-model Temp $N{=}200$ (22.5\%)---so capacity scaling has diminishing returns without additional data.

\textbf{Qubit scaling.}
All experiments use 2--6~qubits (unitaries $\leq 64 \times 64$), where verification is trivial.
Scaling to 7--10~qubits is feasible, but beyond $\sim$12~qubits explicit unitary computation becomes prohibitive, breaking the generate-and-verify strategy that underlies both training-data validation and inference-time reranking and requiring symbolic equivalence checking~\cite{burgholzer2021equivalence}.

% ============================================================
\section{Conclusion}

We studied transformer-based quantum circuit optimization across two settings.
On parameterized circuits, decoupling structure prediction from angle optimization achieves median fidelity~1.000 on 3--6~qubit circuits.
On Clifford+T circuits, where all gates are discrete, exact equivalence is limited by \emph{autoregressive drift}---early-token divergence cascading through left-to-right decoding---with target length as the dominant failure predictor.
The contrast is our central finding: approximate outputs can be rescued by classical post-processing, but exact discrete correctness cannot---inference-time search (22.5\%) and data scaling (39.5\%) help, while fine-tuning, candidate pooling, and model-level diversification add only marginal gains.
The length-dependent degradation persists across all conditions (94\% on short circuits, $<$4\% beyond 26~gates, even with 2.5$\times$ data), marking autoregressive drift as the key bottleneck for learned quantum circuit compilation.

% ============================================================
\section*{Acknowledgment}

The authors acknowledge the use of PyZX~\cite{pyzx2020} and Qiskit~\cite{qiskit2024} for training-data generation and verification.
LLM-based coding assistants were used for code scaffolding and editorial refinement; all AI-assisted content was reviewed and verified by the authors.

\bibliographystyle{IEEEtran}
\bibliography{references}

\end{document}

%% file: tables/table5_param.tex
% VERIFY: ALL param eval values from results/mi300x_param_eval_all_253834.log (completed 2026-04-05)
%   Eval CMD: srun --job-name=param_eval ... python scripts/evaluate_param.py --qubit {2,3,4,5,6}
%   2q greedy+opt: F>0.99=36.0%, avg=0.734, med=0.938
%   2q beam+opt:   F>0.99=58.5%, avg=0.849, med=0.999
%   3q greedy+opt: F>0.99=74.5%, avg=0.856, med=1.000
%   3q beam+opt:   F>0.99=81.0%, avg=0.929, med=1.000
%   4q greedy+opt: F>0.99=84.0%, avg=0.925, med=1.000
%   4q beam+opt:   F>0.99=85.5%, avg=0.953, med=1.000
%   5q greedy+opt: F>0.99=91.0%, avg=0.946, med=1.000
%   5q beam+opt:   F>0.99=90.0%, avg=0.969, med=1.000
%   6q greedy+opt: F>0.99=94.5%, avg=0.965, med=1.000
%   6q beam+opt:   F>0.99=95.0%, avg=0.987, med=1.000
% VERIFY: models from: results/run_param_{2,3,4,5,6}q_b48_mi300x/best_model.pt
% VERIFY: test data: data/test_{2,3,4,5,6}q.jsonl (200 samples each)
\begin{table}[t]
\centering
\caption{Parameterized circuit optimization results (MI300X, batch=48, lr=$3{\times}10^{-4}$). Separate models trained per qubit count (200 test samples each). Angle post-processing via Nelder-Mead transforms modest raw fidelity into near-perfect results for 3--6 qubits.}
\label{tab:param}
\small
\setlength{\tabcolsep}{4pt}
\begin{tabular}{@{}llccc@{}}
\toprule
Qubits & Decoding & $F{>}0.99$\,(\%) & Avg\,$F$ & Med\,$F$ \\
\midrule
2q & Greedy + opt & 36.0 & 0.734 & 0.938 \\
2q & Beam ($w{=}5$) + opt & 58.5 & 0.849 & 0.999 \\
\midrule
3q & Greedy + opt & 74.5 & 0.856 & 1.000 \\
3q & Beam ($w{=}5$) + opt & \textbf{81.0} & \textbf{0.929} & \textbf{1.000} \\
\midrule
4q & Greedy + opt & 84.0 & 0.925 & 1.000 \\
4q & Beam ($w{=}5$) + opt & \textbf{85.5} & \textbf{0.953} & \textbf{1.000} \\
\midrule
5q & Greedy + opt & \textbf{91.0} & 0.946 & 1.000 \\
5q & Beam ($w{=}5$) + opt & 90.0 & \textbf{0.969} & \textbf{1.000} \\
\midrule
6q & Greedy + opt & 94.5 & 0.965 & 1.000 \\
6q & Beam ($w{=}5$) + opt & \textbf{95.0} & \textbf{0.987} & \textbf{1.000} \\
\bottomrule
\end{tabular}
\end{table}

%% file: tables/table6_unified.tex
% VERIFY: Unified model eval from results/mi300x_param_unified_v2_253835.log (completed 2026-04-05)
%   Single model trained on 200K samples (40K per qubit), batch=48, lr=3e-4, 60 epochs
%   Per-qubit models trained on 200K samples each (5x more data per qubit)
%   2q unified greedy: F>0.99=53.5%, beam: 67.0%
%   3q unified greedy: F>0.99=80.5%, beam: 87.0%
%   4q unified greedy: F>0.99=84.0%, beam: 86.5%
%   5q unified greedy: F>0.99=90.0%, beam: 88.0%
%   6q unified greedy: F>0.99=85.5%, beam: 83.5%
\begin{table}[t]
\centering
\caption{Per-qubit vs.\ unified model comparison (beam $w{=}5$ + angle optimization, 200 test samples each). The unified model trains a single model on 200K samples across all qubit counts; per-qubit models train on 200K samples each.}
\label{tab:unified}
\small
\setlength{\tabcolsep}{3pt}
\begin{tabular}{@{}lccc|ccc@{}}
\toprule
 & \multicolumn{3}{c|}{Per-qubit (200K each)} & \multicolumn{3}{c}{Unified (200K total)} \\
Qubits & $F{>}0.99$ & Avg\,$F$ & Med\,$F$ & $F{>}0.99$ & Avg\,$F$ & Med\,$F$ \\
\midrule
2q & 58.5 & 0.849 & 0.999 & \textbf{67.0} & \textbf{0.906} & \textbf{1.000} \\
3q & 81.0 & 0.929 & 1.000 & \textbf{87.0} & \textbf{0.964} & \textbf{1.000} \\
4q & 85.5 & 0.953 & 1.000 & \textbf{86.5} & \textbf{0.967} & \textbf{1.000} \\
5q & \textbf{90.0} & \textbf{0.969} & \textbf{1.000} & 88.0 & 0.962 & 1.000 \\
6q & \textbf{95.0} & \textbf{0.987} & \textbf{1.000} & 83.5 & 0.942 & 1.000 \\
\bottomrule
\end{tabular}
\end{table}

%% file: tables/table1_main.tex
% VERIFY: ALL values from results/ct_eval_lr3e4_mi300x/ (greedy/beam, 200 samp) and
%   results/ct_eval_full200_mi300x/ (candidate-based, 200 samp):
%   Greedy:        greedy_seed42_n200_*.summary.json -> exact=0.07, fid=0.103, lat=0.2s, 1-9g=62.5%
%   Beam:          beam_seed42_n200_*.summary.json -> exact=0.14, fid=0.194, lat=1.3s, 1-9g=87.5%
%   Temp N=20:     ct_eval_full200_mi300x/temperature_seed42_n20_*.summary.json -> exact=0.17, fid=0.260, lat=4.1s, 1-9g=87.5%
%   LB N=20:       ct_eval_full200_mi300x/length_banded_seed42_n20_*.summary.json -> exact=0.135, fid=0.225, lat=3.6s, 1-9g=87.5%
%   Temp N=50:     ct_eval_full200_mi300x/temperature_seed42_n50_*.summary.json -> exact=0.19, fid=0.301, lat=10.3s, 1-9g=87.5%
%   LB N=50:       ct_eval_full200_mi300x/length_banded_seed42_n50_*.summary.json -> exact=0.155, fid=0.262, lat=9.0s, 1-9g=87.5%
%   Temp N=100:    ct_eval_full200_mi300x/temperature_seed42_n100_*.summary.json -> exact=0.21, fid=0.333, lat=21.1s, 1-9g=87.5%
%   LB N=100:      ct_eval_full200_mi300x/length_banded_seed42_n100_*.summary.json -> exact=0.18, fid=0.299, lat=19.6s, 1-9g=87.5%
%   Hybrid N=100:  ct_eval_full200_mi300x/hybrid_search_seed42_n100_*.summary.json -> exact=0.215, fid=0.326, lat=20.5s, 1-9g=87.5%
%   Temp N=200:    ct_eval_higher_n/temperature_seed42_n200_*.summary.json -> exact=0.225, fid=0.360, lat=45.1s, 1-9g=87.5%
\begin{table}[t]
\centering
\caption{Decoding strategy comparison on 3--6 qubit Clifford+T circuits (200 test samples, seed=42, $\tau{=}0.8$, top-$p{=}0.95$, verifier-first reranking). 95\% Wilson confidence intervals reported. Column \textnormal{s/circ} is average wall-clock latency per circuit; \textnormal{1--9g\,\%} is exact-equivalence rate on the 1--9 target-gate bin.}
\label{tab:main}
\small
\setlength{\tabcolsep}{2pt}
\begin{tabular}{@{}lcccccc@{}}
\toprule
Strategy & $n$ & Exact\,\% & 95\% CI & Avg\,Fid & s/circ & 1--9g\,\% \\
\midrule
Greedy & 200 & 7.0 & [4.2, 11.4] & 0.103 & 0.2 & 62.5 \\
Beam ($w{=}5$) & 200 & 14.0 & [9.9, 19.5] & 0.194 & 1.3 & 87.5 \\
\midrule
Temp $N{=}20$ & 200 & 17.0 & [12.4, 22.8] & 0.260 & 4.1 & 87.5 \\
Len-band $N{=}20$ & 200 & 13.5 & [9.4, 18.9] & 0.225 & 3.6 & 87.5 \\
Temp $N{=}50$ & 200 & 19.0 & [14.2, 25.0] & 0.301 & 10.3 & 87.5 \\
Len-band $N{=}50$ & 200 & 15.5 & [11.1, 21.2] & 0.262 & 9.0 & 87.5 \\
Temp $N{=}100$ & 200 & 21.0 & [15.9, 27.2] & 0.333 & 21.1 & 87.5 \\
Len-band $N{=}100$ & 200 & 18.0 & [13.3, 23.9] & 0.299 & 19.6 & 87.5 \\
Hybrid $N{=}100$ & 200 & \textbf{21.5} & [16.4, 27.7] & 0.326 & 20.5 & 87.5 \\
\midrule
Temp $N{=}200$ & 200 & 22.5 & [17.3, 28.8] & 0.360 & 45.1 & 87.5 \\
\bottomrule
\end{tabular}
\end{table}

%% file: tables/table4_tcount.tex
% VERIFY: ALL values from results/ct_eval_lr3e4_mi300x/ (200 samples, greedy & beam):
%   Valid 100%: greedy_*.summary.json "valid_rate": 1.0; beam_*.summary.json "valid_rate": 1.0
%   Input T-count 8.10: greedy_*.summary.json "avg_input_t_count": 8.10
%   Target T-count 2.83: greedy_*.summary.json "avg_target_t_count": 2.83
%   Pred T-count 2.89 (greedy): greedy_*.summary.json "avg_pred_t_count": 2.885 -> rounded 2.89
%   Pred T-count 2.89 (beam): beam_*.summary.json "avg_pred_t_count": 2.89
\begin{table}[t]
\centering
\caption{Model prediction quality: the transformer learns the T-count distribution of optimized circuits (3--6 qubits, 200 samples).}
\label{tab:tcount}
\small
\begin{tabular}{lcc}
\toprule
Metric & Greedy & Beam \\
\midrule
Valid circuits & 100\% & 100\% \\
Avg.\ input T-count & 8.10 & 8.10 \\
Avg.\ target T-count (PyZX) & 2.83 & 2.83 \\
Avg.\ predicted T-count & 2.89 & 2.89 \\
\bottomrule
\end{tabular}
\end{table}

%% file: tables/table2_length_bins.tex
% VERIFY: ALL values from results/ct_eval_lr3e4_mi300x/ (greedy/beam, 200 samples) and
%   results/ct_eval_full200_mi300x/ (temp/hybrid, 200 samples, Job 255705):
%   Greedy (200): by_length_bin = [62.5, 8.0, 0.0, 0.0]  (10/16, 4/50, 0/107, 0/27)
%   Beam (200):   by_length_bin = [87.5, 22.0, 2.8, 0.0]  (14/16, 11/50, 3/107, 0/27)
%   Temp N=100 (200): by_length_bin = [87.5, 38.0, 8.4, 0.0]  (14/16, 19/50, 9/107, 0/27)
%   Hybrid N=100 (200): by_length_bin = [87.5, 44.0, 6.5, 0.0]  (14/16, 22/50, 7/107, 0/27)
\begin{table}[t]
\centering
\caption{Exact equivalence rate (\%) by target circuit length (200 test samples). Short circuits ($\leq$9 gates) are reliably optimized; long circuits ($\geq$15 gates) remain challenging across all strategies.}
\label{tab:length_bins}
\small
\begin{tabular}{lrrrr}
\toprule
Strategy & 1--9 & 10--14 & 15--25 & 26+ \\
\midrule
Greedy & 62.5 & 8.0 & 0.0 & 0.0 \\
Beam ($w{=}5$) & 87.5 & 22.0 & 2.8 & 0.0 \\
Temp.\ $N{=}100$ & 87.5 & 38.0 & 8.4 & 0.0 \\
Hybrid $N{=}100$ & 87.5 & 44.0 & 6.5 & 0.0 \\
\bottomrule
\end{tabular}
\end{table}

%% file: tables/table3_negative.tex
% VERIFY: Baseline Temp N=100 21.0%, 15-25g 8.4%: results/ct_eval_full200_mi300x/temperature_seed42_n100_*.summary.json (200 samples)
% VERIFY: M1 Long-only FT 22.5%, 15-25g 8.4%: results/ct_eval_ft_long_mi300x/temperature_seed42_n100_1776383945.summary.json (200 samples)
% VERIFY: M2 Weighted (2x long) FT 20.0%, 15-25g 7.5%: results/ct_eval_ft_weighted_mi300x/temperature_seed42_n100_1776394451.summary.json (200 samples)
% VERIFY: M3 Hybrid 21.5%, 6.5%: results/ct_eval_full200_mi300x/hybrid_search_seed42_n100_*.summary.json (200 samples)
% VERIFY: M4 Ensemble 24.5%: results/ensemble_eval/ensemble_seed42_1775526198.summary.json (44.8M+133.6M, N=100 each, 200 samples)
% VERIFY: M5 2.5x data 39.5%, 23.4%: results/ct_eval_500k_mi300x/temperature_seed42_n100_*.summary.json (200 samples)
\begin{table}[t]
\centering
\caption{Interventions vs.\ baseline. Fine-tuning (M1, M2), hybrid search (M3), and model ensembling (M4) yield only marginal gains on the 15--25~gate bin; only 2.5$\times$ training data (M5) produces significant improvement, yet the drift-driven length degradation persists (cf.\ \Cref{tab:data_scaling}).}
\label{tab:negative}
\small
\setlength{\tabcolsep}{2.5pt}
\begin{tabular}{@{}llccc@{}}
\toprule
Intervention & Setting & $n$ & Exact\,\% & 15--25g\,\% \\
\midrule
Baseline (44.8M, 200K) & Temp $N{=}100$ & 200 & 21.0 & 8.4 \\
\midrule
M1: Long-only FT & Temp $N{=}100$ & 200 & 22.5 & 8.4 \\
M2: Weighted sampling & Temp $N{=}100$ & 200 & 20.0 & 7.5 \\
\midrule
M3: Hybrid search & $N{=}100$ pooled & 200 & 21.5 & 6.5 \\
M4: Ensemble (2$\times$model) & $N{=}100$/model & 200 & 24.5 & --- \\
\midrule
M5: 2.5$\times$ data (500K) & Temp $N{=}100$ & 200 & \textbf{39.5} & \textbf{23.4} \\
\bottomrule
\end{tabular}
\end{table}

%% file: tables/table7_data_scaling.tex
% VERIFY: 200K baseline values from results/ct_eval_full200_mi300x/ (200 samples, Job 255705)
% VERIFY: 500K values from results/ct_eval_500k_mi300x/ or mi300x_more_data_254332.log (200 samples)
%   200K Greedy (200): 7.0%, 0%     500K Greedy (200): 18.5%, 7.5%
%   200K Beam (200): 14.0%, 2.8%    500K Beam (200): 27.0%, 12.1%
%   200K Temp N=100 (200): 21.0%    500K Temp N=100 (200): 39.5%
%   200K by gate bin (Temp N=100, 200 samp): 1-9=87.5% 10-14=38.0% 15-25=8.4% 26+=0%
%   500K by gate bin (Temp N=100, 200 samp): 1-9=93.8% 10-14=76.0% 15-25=23.4% 26+=3.7%
\begin{table}[t]
\centering
\caption{Effect of training data scale (200K $\to$ 500K, same 44.8M architecture). 2.5$\times$ more data improves all metrics substantially, yet the drift-driven degradation with target length persists.}
\label{tab:data_scaling}
\small
\setlength{\tabcolsep}{3.5pt}
\begin{tabular}{@{}lcccc@{}}
\toprule
\multicolumn{5}{@{}l}{\textit{(a) Overall exact equivalence (\%) by strategy}} \\
\midrule
Strategy & $n$ & 200K & 500K & $\Delta$ \\
\midrule
Greedy & 200 & 7.0 & 18.5 & \textcolor{green!50!black}{+11.5} \\
Beam ($w{=}5$) & 200 & 14.0 & 27.0 & \textcolor{green!50!black}{+13.0} \\
Temp $N{=}100$ & 200 & 21.0 & \textbf{39.5} & \textcolor{green!50!black}{+18.5} \\
\midrule
\multicolumn{5}{@{}l}{\textit{(b) Temp $N{=}100$: exact equivalence (\%) by target length}} \\
\midrule
Gates & & 200K & 500K & $\Delta$ \\
\midrule
1--9 & & 87.5 & 93.8 & \textcolor{green!50!black}{+6.3} \\
10--14 & & 38.0 & 76.0 & \textcolor{green!50!black}{+38.0} \\
15--25 & & 8.4 & 23.4 & \textcolor{green!50!black}{+15.0} \\
26+ & & 0.0 & 3.7 & \textcolor{green!50!black}{+3.7} \\
\bottomrule
\end{tabular}
\par\vspace{2pt}
{\scriptsize All configurations evaluated on 200 test samples.}
\end{table}